\documentstyle[11pt,newpasp,twoside,epsf,psfig]{article}
\def\b#1{{\bf #1}}\def\d{{\rm d}}\def\e{{\rm e}}\def\m{\,{\rm
m}}\def\nm{\,{\rm nm}}\def\s{\,{\rm s}}
\def\pc{\,{\rm pc}}\def\kpc{\,{\rm kpc}}\def\Mpc{\,{\rm Mpc}}
\def\kms{\,{\rm km}\,{\rm s^{-1}}}\def\msun{\,{\rm M}_\odot}
\begin{document}
\title{Microlensing and Galactic Structure}
 \author{James Binney}
\affil{Oxford University, Theoretical Physics, Keble Road, Oxford, OX1 3NP,
U.K.}

\begin{abstract}
Because we know little about the Galactic force-field away from the plane,
the Galactic mass distribution is very ill-determined. I show that a
microlensing survey of galaxies closer than $50\Mpc$ would enable us to map
in three dimensions the Galactic density of stellar mass, which should be strictly
less than the total mass density. A lower limit can be placed on the stellar
mass needed at $R<R_0$ to generate the measured optical depth towards
sources in the bulge. If the Galaxy is barred, this limit is lower by a
factor of up to two than in the axisymmetric case.  Even our limited knowledge
of the Galactic force field suffices to rule out the presence of the amount
of mass an axisymmetric Galaxy needs to generate the measured optical depth.
Several lines of argument imply that the Galaxy is strongly barred only at
$R<4\kpc$, and if this is the case, even barred Galaxy models cannot
generate the measured optical depth without violating some constraint on the
Galactic force-field. Galactic mass models that are based on the assumption
that light traces mass, for which there is significant support in the inner
Galaxy, yield microlensing optical depths that are smaller than the measured
value by a factor of more than 2.5.
\end{abstract}

\section{Introduction}

Notwithstanding the difficulties to which heavy obscuration by dust gives
rise, we now have a reasonable idea of how luminosity is distributed in the
Galaxy. We know we live in a galaxy that has a bar, $3-4\kpc$ long, whose
nearer end lies at positive longitudes. Within the bar there is a moderately
flattened component in which the luminosity density rises roughly as
$r^{-1.8}$ with decreasing galactocentric radius $r$. Around the bar there
is a roughly exponential disk with scale length $\sim3\kpc$, and a
moderately flattened metal-poor halo in which the luminosity density varies
as $\sim r^{-3.5}$.

The situation regarding the Galactic distribution of mass is very much less
satisfactory. Observations of external galaxies and cosmological theory have
convinced us that light is not a good tracer of mass. Hence we cannot
straightforwardly translate our models of the luminosity distribution into
models of the mass distribution; in principle, we should start afresh and
derive the mass distribution by tracing out the Galaxy's gravitational field
$\b F(\b r)$ and then applying the divergence operator: $\nabla\cdot\b F=-4\pi
G\rho$.

From observations of gas that flows in the Galactic plane we have a fair
idea of what $\b F(\b r)$ looks like at points $\b r$ within the plane, but
we have very little secure knowledge of $\b F$ out of the plane. This
deficiency is serious, because until we know $\b F$ away from the plane, we
cannot apply $\nabla$ {\em anywhere}, and can say {\em nothing\/} with
security about $\rho$ {\em anywhere\/}!  

What we currently know about $\b F$ out of the plane relates to points near
the Sun. Cr\'ez\'e et al.\ (1998) and Holmberg \& Flynn (2000) have used
proper motions of Hipparcos stars to estimate the gradient of $F_z$ near the
plane and hence infer that the local mass density is
$0.076\pm0.015\msun\pc^{-3}$ and $0.10\pm0.01\msun\pc^{-3}$, respectively,
while Kuijken \& Gilmore (1991) have used the radial velocities of stars at
the SGP to estimate $F_z$ at $z=-1.1\kpc$ and thus infer that the surface
density of the Galaxy within $1.1\kpc$ of the plane is $71\pm6\msun\pc^{-3}$. To
constrain $\b F$ at points that lie far from the Sun, we have to study
objects that move to such points.  High-velocity stars are the obvious
tracers to use, because they are so numerous. They strongly constrain models
of $\b F$ because stars on essentially the same orbit can be studied both
locally and in situ (Binney, 1994; Dehnen \& Binney, 1996). Unfortunately, the potential of
this approach has yet to be systematically exploited. Recently, there has
been considerable interest in using tidal streamers associated with
disrupted satellites as tracers of $\b F$ (Johnston et al.\ 1999, Helmi et
al.\ 1999). My own view is that high-velocity stars have greater potential
because (a) they are vastly more numerous and (b) they do not require
approximations of the level of the (manifestly false) assumption that all
elements of a streamer are on the same orbit.  Moreover, extracting useful
information from streamers requires space-based astrometry, the requisite
observations high-velocity stars are available now, and all that's lacking
is machinery for modelling them.

Gravitational microlensing directly probes the Galaxy's mass distribution,
but in a very different way from classical studies of gas and stars. In
fact, microlensing does not measure the smooth Galactic force-field $\b F$
but  graininess in the mass distribution. Hence, it is insensitive to the
contribution to the latter from elementary particles and gas, and is therefore
complementary to the traditional approach to the determination of $\rho$
from $\b F$, rather than competitive with it.

\section{From optical depth to stellar density}

The optical depth to microlensing of a stellar object at distance $s_0$ is
 \begin{equation}\label{basiceq}
\tau={4\pi G\over c^2}\int_0^{s_0}\d s\,\rho_*\hat s,
\end{equation}
where
\begin{equation}
\hat s=\Big({1\over s_0-s}+{1\over s}\Big)^{-1}
\end{equation}
 is the harmonic mean of the source-lens and lens-observer distances.
If the source is extragalactic, $\hat s\simeq s$, and $\tau$ becomes
proportional to $\int\d s\,\rho_* s$ in the direction of the source. Suppose
we measure $\tau$ along many lines of sight all over the sky. Can we
reconstruct $\rho_*$ from the data? 

\begin{figure}
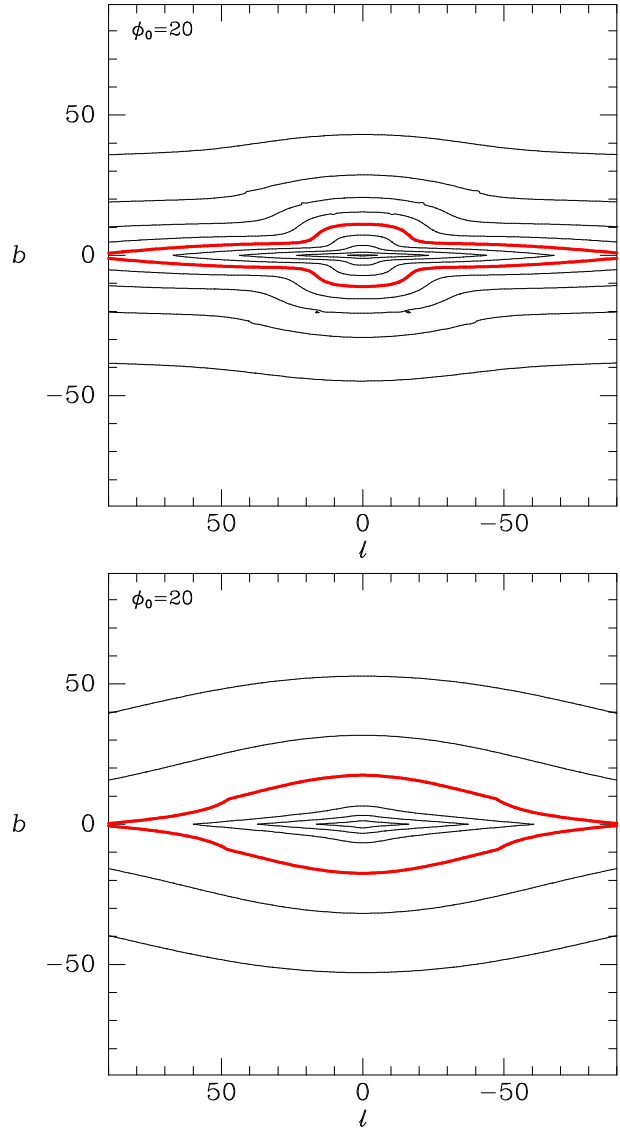

\centerline{\psfig{file=kent.ps,width=.6\hsize}}
\vskip5pt
\centerline{\psfig{file=dehnen.ps,width=.6\hsize}}
\caption{Contours of equal microlensing optical depth for a distant source
for two Galactic models. There are two contours per decade and the heavy
contour is for $\tau=10^{-6}$. The upper panel is for the model one obtains
from the luminosity model of Kent, Dame \& Fazio (1991) with an assumed
mass-to-light ratio $\Upsilon_K=1$. The lower panel is for Model 1 of Dehnen
\& Binney (1998), and all components, including the dark halo, have been
assumed to contribute fully to $\tau$.  Consequently, the optical depth at
large $|b|$ is larger in the lower than in the upper panel.}
\label{taufig}
\end{figure}

With some simplifying assumptions, we can. Binney \& Gerhard (1996) show
that if the Galactic luminosity density $j(\b r)$ is symmetric about the
Galactic plane and two other, orthogonal planes (as it would be if it were
triaxially ellipsoidal), then a Richardson--Lucy algorithm can be used to
recover $j(\b r)$ from its line-of-sight projection, $I(\Omega)=\int\d s\,j$, at
each point $\Omega$ on the celestial sphere.  It is straightforward to
modify the derivation of Binney \& Gerhard to show that, with the same
assumptions regarding symmetry, a Richardson--Lucy algorithm for the
recovery of $\rho_*$ from $\tau(\Omega)$ is
 \begin{equation}\label{LReq}
\rho_*^{(k+1)}(\b r)=\rho_*^{(k)}(\b r){{\displaystyle\sum_{i=1}^8
{\tau(\Omega_i)\over\tau^{(k)}(\Omega_i)}{1\over s(\b r_i)}}\bigg/
{\displaystyle\sum_{i=1}^8{1\over s(\b r_i)}}}.
\end{equation}
 Here the sum over $i$ is over the eight points $\b r_i$ that are connected
to $\b r$ by the assumed symmetry of $\rho_*$; the lines of sight to these
points are in the directions $\Omega_i$, and their distances from the Sun
are $s(\b r_i)$.

While it is in principle possible to recover $\rho_*$ from $\tau$, it is not
clear that this will ever be done. The problem is the small numerical factor
in front of the integral in (1). Numerically,
 \begin{equation}\label{dimlesseq}
\tau=6\times10^{-6}\left[
{\int_0^{s_0}\d s\,\rho_* s\over 10^{10}\msun/\!\kpc}\right],
\end{equation}
 where we have again assumed that the source is extragalactic.  Figure
\ref{taufig} plots $\tau$ for two typical Galactic models. The model
underlying the upper panel is the $K$-band luminosity model of Kent, Dame \&
Fazio (1991), which has been converted into a model of the stellar mass
distribution by assuming a mass-to-light ratio $\Upsilon_K=1$, which appears
to be correct for the solar neighbourhood (\S 10.4.4 of Binney \& Merrifield
1998). One sees that $\tau$ exceeds $10^{-6}$ only for lines of sight at
fairly low latitudes.

Existing data indicate that the duration of a microlensing event is
typically tens of days, so in an observing season one obtains at most a
handful of statistically independent observations per line of sight. Hence,
in a given area of the sky the number of lines of sight that must be
monitored for of order years to distinguish $\tau$ from zero is
$\sim1/\tau\ga10^6$. Finding this number of extragalactic sources in each of
a large number of patches of the sky is hard. Probably our best chance is
offered by `pixel lensing' towards nearby galaxies. Gould (1996) gives the
necessary theory and I adopt his notation. A large r\^ole is played in this
by the effective stellar flux $F_*$, which is defined in terms of the
stellar luminosity function $\phi$ by
 \begin{equation}\label{defsFstar}
F_*=\int\d F\,\phi(F)F^2\bigg/\int\d F\,\phi(F)F.
\end{equation}
 Per resolution element on a galaxy image, the rate  at which microlensing
events can be detected above a signal-to-noise threshold $Q_{\rm min}$ is
 \begin{equation}\label{givesrate}
\Gamma={2\kappa\xi\over Q_{\rm min}^2}\tau\alpha F_*,
\end{equation}
 where $1>\kappa,\xi$ are dimensionless functions, and $\alpha F_*$ is the
rate at which the telescope detects photons from an object of flux $F_*$. In
terms of the surface brightnesses of galaxy and sky $S$ and $S_{\rm sky}$,
we have $\kappa\equiv (1+S_{\rm sky}/S)^{-1}\simeq S/S_{\rm sky}$ over most
of a galactic image. The value of $\xi$ depends on the degree to which the
image is resolved into stars: if a significant part of the integral on the
top of equation (\ref{defsFstar}) comes from stars bright enough that
lensing of them can be detected even at large impact parameter
($u\ga0.25$), then $\xi$ is small, with $\xi\sim1$ otherwise.

Gould shows that if one is interested only in measuring $\tau$ regardless of
the masses of the lenses that generate it, the optimal observational
strategy is to work in the regime $\xi\sim1$ in which one detects only
high-magnification events. We shall be in this regime provided
\begin{equation}\label{spikecond}
{\kappa\alpha F_*^2\over\omega F_{\rm psf}}<{Q_{\rm min}^2\over4\pi},
\end{equation}
 where $\omega$ is the rate constant of a typical microlensing event and
$F_{\rm psf}$ is the flux in a resolution element of the galactic image. How
big will these quantities be in a typical case? Suppose we are using a
diffraction-limited telescope of diameter $D$ to study a galaxy of distance
$s$ at radius $R_{25}$, where the $V$-band surface brightness will be
$\sim24\,\hbox{mag}\,\hbox{arcsec}^{-2}$. Gould gives the absolute magnitude
corresponding to $F_*$ as
\begin{equation}\label{defsIstar}
M_{*I}=-4.84+3(V-I)
\end{equation}
and estimates that from a star with $I=20$ our telescope will collect
photons at a rate $10(D/1\m)^2\s^{-1}$. Hence,
\begin{equation}\label{givesalphaF}
\alpha F_*=10\Big({D\over1\m}\Big)^2 10^{-0.4(30-20-4.84+3(V-I))}
\Big({s\over10\Mpc}\Big)^{-2}.
\end{equation}
The $I$-band flux in the telescope's resolution element is
\begin{equation}\label{givesFpsf}
F_{\rm psf}=F_010^{-0.4(24-(V-I))}\Big[0.206\Big({\lambda/1000\nm\over D/1\m}\Big)\Big]^2
,
\end{equation}
 where $F_0$ is some universal constant. We can express $F_*$ in terms of
 this same constant and $M_{*I}$ thus
\begin{equation}\label{givesFstar}
F_*=F_0 10^{-0.4(30-4.84+3(V-I))}\Big({s\over10\Mpc}\Big)^{-2}.
\end{equation}
 When we substitute equations (\ref{givesalphaF}), (\ref{givesFpsf}) and
(\ref{givesFstar}) into equation (\ref{spikecond}) and assume (Table 4.4 of
Binney \& Merrifield, 1998)
\begin{equation}\label{givesK}
\kappa\sim10^{-0.4(4-(V-I))}
\end{equation}
 and $V-I\sim1.25$ (de Jong, 1995) the condition to 
be in the high-magnification regime becomes
\begin{equation}
10.4
\Big({D\over1\m}\Big)^4
\Big({s\over10\Mpc}\Big)^{-4}\Big({\lambda\over1000\nm}\Big)^{-2}\big(\omega1\hbox{week}\big)^{-1}
<{Q_{\rm min}^2\over4\pi}.
\end{equation}
Gould finds that $Q_{\rm min}\sim7$ is required for detection, so this
condition is satisfied for $D\la4\m$ and $s\sim50\Mpc$.
By equations (\ref{givesrate}), (\ref{givesalphaF}) and (\ref{givesK}) 
the event rate in this regime is
\begin{eqnarray}
\Gamma&=&{2\tau\over Q_{\rm min}^2}10^{-0.4(6.66+2(V-I))}
\Big({D\over1\m}\Big)^2\Big({s\over10\Mpc}\Big)^{-2}\s^{-1} \nonumber\\
&=&{262\tau\over Q_{\rm min}^2}\Big({D\over1\m}\Big)^2\Big({s\over10\Mpc}\Big)^{-2}\,\hbox{week}^{-1}
\end{eqnarray}
 Hence, by directing a $4\m$ telescope along a line of sight with
$\tau=10^{-6}$, we should be able to detect three to four events per $10^6$
resolution elements per week. In fact a somewhat higher event rate could be
achieved if the telescope were in space, because the event rate is inversely
proportional to the sky brightness, which will at least $2\,$mag fainter in
space than the ground-based value I have assumed.
 
If the PSF of the telescope has FWHM of $x$ arcsec, a galaxy of diameter $y$
arcmin offers 3600$(y/x)^2$ pixels to monitor. An $L_*$ galaxy at a distance
of $50\Mpc$ has $y\sim1$, and will provide $10^6$ pixels if $x\sim0.06$.
Thus, a diffraction-limited telescope of modest aperture monitoring galaxies
within $50\Mpc$ for of order a year could map $\tau$ sufficiently
extensively for it to be possible to recover $\rho_*$ from a scheme such as
(\ref{LReq}).

Of course, the values of $\tau$ recovered from such a survey would include
contributions from self-lensing within the target object in addition to the
optical depth through the Milky Way. As Crotts (1992) pointed out in
relation to pixel lensing of M31, variations in $\tau$ across the image of
an highly inclined galaxy would help one to determine how much self lensing
occurs in galactic halos rather than in disks or bulges. Consequently, the
analysis of the data from the survey would proceed by modelling in some
detail the distribution of $\tau$ in each target galaxy, with the Milky
Way's contribution in that direction as a single number to be fitted to a
considerable body of data.

\section{Real data}

Before we lobby NASA and ESA for a dedicated microlensing space telescope,
we should ask what can be learned from the existing microlensing data. Three
areas of the sky  have been monitored: (i) towards the Galactic bulge; (ii)
towards some spiral arms; and (iii) towards the Magellanic Clouds.

If one believes that the dark halo is comprised of elementary particles,
only a very small optical depth is predicted towards the Clouds
($\sim4\times10^{-8}$). At $1.2^{+0.4}_{-0.3}\times10^{-7}$ the measured
optical depth is about three times larger, but still deriving from only 17
events (Alcock et al.\ 2000a). The nature of these events is controversial.
There is a powerful case that many of the lenses lie in the Clouds
themselves (Kerins \& Evans, 1999) because the only two lenses with
reasonably securely determined distances (from finite-source effects) do lie
in the Clouds. On the other hand, Ibata et al.\ (1999) may have detected a
substantial population of old white dwarfs from their proper motions in the
HDF. It is just possible that these objects provide the lenses for a
significant number of the observed events. The arguments against a large
column density in white dwarfs remain powerful, however (Gibson \& Mould
1997).

The EROS collaboration (2000) has observed seven probable microlensing
events in the directions of spiral arms, and from them derived an optical
depth $4.5^{+2.4}_{-1.1}\times10^{-7}$ that is compatible with conventional
Galaxy models. Unfortunately, the error bar on this measurement is rather large.

\subsection{The Galactic centre}

Several hundred events have been detected by various groups along
lines of sight towards the Galactic centre, and these data pose a fascinating
puzzle. They are harder to interpret than the data for lines of sight to the
Clouds because we do not know a priori where the sources are. However, if we
confine ourselves to the data for red-clump stars, we can have quite a
precise estimate of their distribution down each line of sight.

The red-clump stars must follow the general distribution of near infra-red
light quite closely, because they are part of the population of evolved
stars that are responsible for most of the Galaxy's near-IR luminosity. The
DIRBE experiment aboard the COBE satellite mapped the Galaxy's IR surface
brightness in several wavebands, and in the far-IR, where emission by dust is
dominant. Spergel, Malhotra \& Blitz (1996) used these data to estimate the effects of
extinction on the near-IR data, and produce maps of what the Galaxy would
look like in the near-IR bands in the absence of extinction.

\begin{figure}
\vskip.4\hsize
\centerline{\epsfysize=.75\hsize \epsfbox{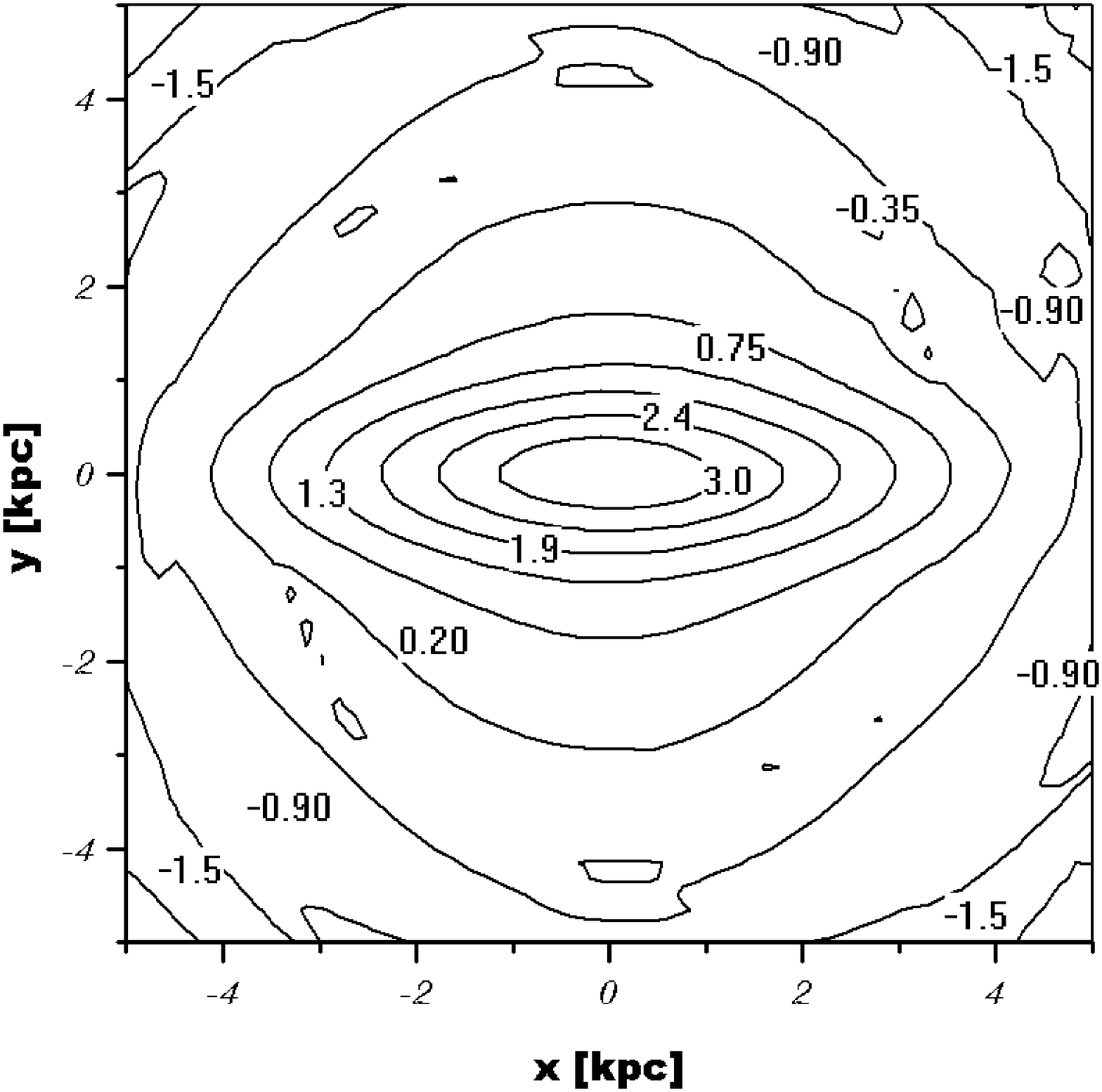}}
\vskip-1.3\hsize
\centerline{\psfig{file=bgs.ps,height=.6\hsize}}
\vskip.65\hsize
\caption{Top panel: the Galaxy in the $L$ band projected perpendicular to
the plane from $z=225\pc$ upwards according to the model of Binney et al.\
(1997). Lower panel: the same view according to the model of Bissantz \&
Gerhard (2000).  In both plots contours are logarithmically spaced. In the
upper plot there are three contours per decade, while below contours are
explicitly labelled.\label{barfig}}
\end{figure}

Binney, Gerhard \& Spergel (1997) used their Richardson--Lucy algorithm to
deproject these corrected near-IR data under the assumption that the Galaxy
has three specified perpendicular planes of mirror symmetry. The upper panel
of Figure \ref{barfig} shows the luminosity density that they recovered when
projected perpendicular to the plane from $z=225\pc$ upwards.  Excluding the
Galactic plane from the projection suppresses local maxima at $\sim3\kpc$
along the $y$ that are a notable feature of the density distribution in the
plane (see Fig.~5 of Binney et al.).  Binney et al argued that these maxima
were artifacts resulting from the presence in the Galactic plane of spiral
arms, which violate the assumed eight-fold symmetry.

Bissantz \& Gerhard (2000) have recently deprojected the same data with an
entirely different technique. Rather than using the Richardson--Lucy
algorithm, they formulate the deprojection problem as a regularized
likelihood maximization. They do not directly  impose any  symmetry on the
model but have a term in the penalty function that discourages
deviations from eight-fold symmetry. Another term in the penalty function
encourages luminosity to lie along the spiral arms delineated by Ortiz \&
Lepine (1993). 

\begin{figure}
\centerline{\epsfysize=.6\hsize \epsfbox{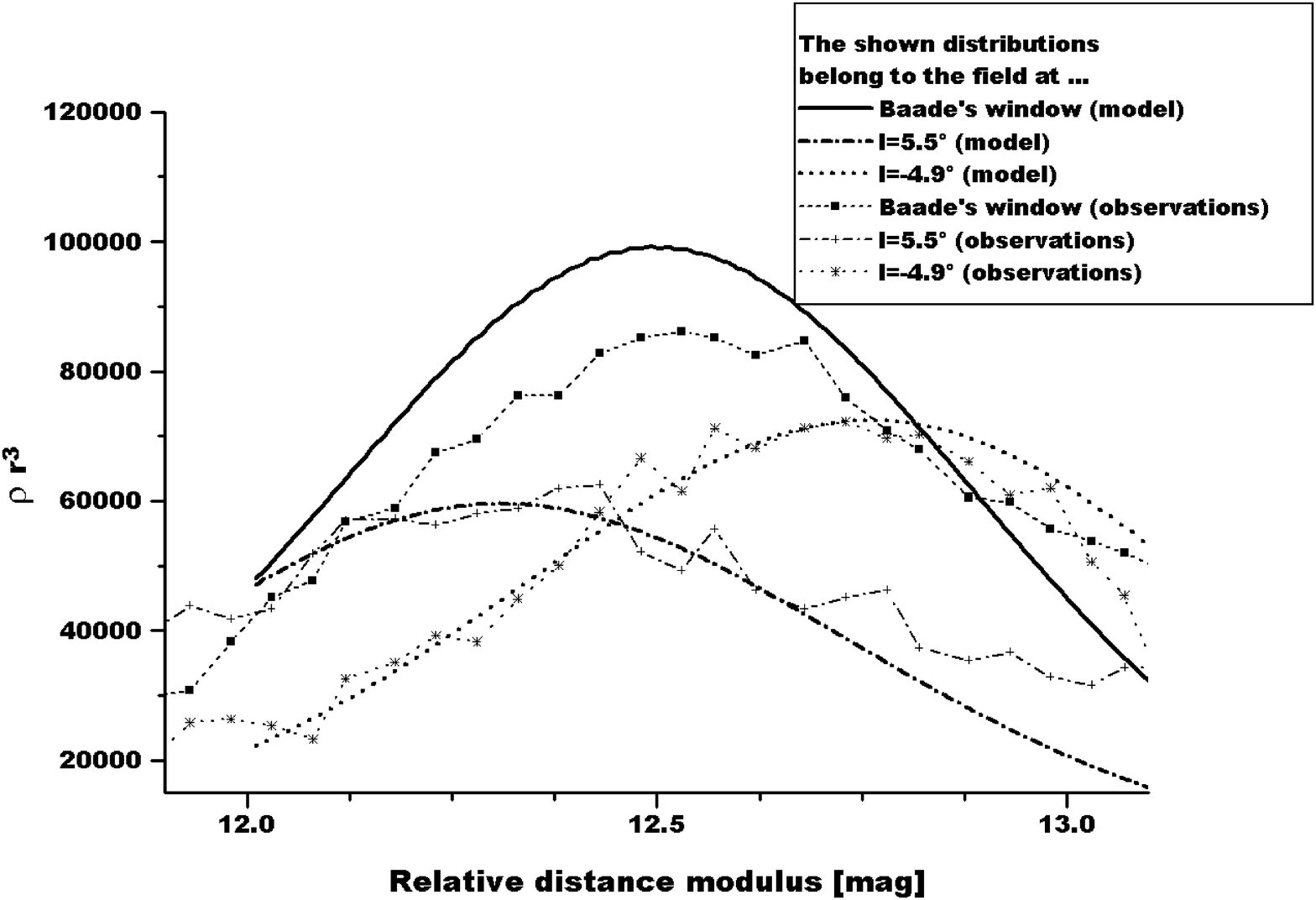}}
\caption{Predicted (curves) and measured (points) apparent-magnitude
distribution of clump stars along three lines of sight through the bulge.
The predictions are based on the model of Bissantz \& Gerhard
(2000).\label{clumpfig}} 
\end{figure}

The lower panel in Figure \ref{barfig} shows that explicitly modelling the
Galaxy's spiral structure in this way has the effect of making the bar
longer and thinner than that recovered by Binney et al.\ (1997) -- the axis
ratio in the plane increases from 2:1 to 3:1. This change to the model bar
enables the latter to reproduce an important datum that the Binney et al bar
did not reproduce: the histograms from Stanek et al.\ (1994) that give for
lines of sight at $l\simeq\pm5\deg$ the number of clump stars at each
apparent magnitude -- see Figure \ref{clumpfig}.  The ability of the model to
reproduce these histograms to good accuracy strongly suggests that the model
gives a faithful account of the distribution of red-clump stars.  Hence,
when this model is used to predict the microlensing optical depth to
red-clump stars, we may be confident that any discrepancy does not arise
from an incorrect distribution of source objects. 

Bissantz \& Gerhard convert their luminosity model of the inner Galaxy into
a mass model by adopting the constant near-IR mass-to-light ratio $\Upsilon$
of Englmaier \& Gerhard (1999). This value of $\Upsilon$ was obtained by
comparing the pattern of gas flow predicted by the Binney et al.\ model for
some assumed $\Upsilon$ to the observed $(l,v)$ diagrams for HI and CO; if
Bissantz \& Gerhard were to repeat this exercise with their new photometric
model, they would surely obtain a very similar value of $\Upsilon$.  Once
$\Upsilon$ has been chosen, one can calculate the microlensing optical depth
for red-clump stars along any line of sight.  In Baade's window they obtain
$\tau=1.24\times10^{-6}$, which is essentially identical to the value
obtained by Bissantz et al.\ (1997) from the shorter, fatter bar of Binney et
al.\ (1997), and significantly short of the values from the MACHO
collaboration: $\tau=3.9^{+1.8}_{-1.2}\times10^{-6}$ ($1\sigma$) directly
measured for bulge clump giants by Alcock et al.\ (1997) and
$(3.9\pm0.6)\times10^{-6}$ estimated by Alcock et al., (2000b) for bulge
stars from a difference-imaging analysis of an inhomogeneous collection of
sources.  Evidently one cannot obtain agreement with the MACHO optical depth
under the assumption that mass follows light.

Since the difference between the optical depth implied by constant
$\Upsilon$ and the measured value is so large, it is natural to investigate
an extreme model in which we ask, what is the minimum mass that is
compatible with the red-clump optical depth attaining the MACHO value? Most
of the red-clump sources lensed in Baade's window $(l,b)=(1\deg,-4\deg)$ lie
close to the Galaxy's $z$ axis, so we simplify the calculation by
considering a source that lies distance $h$ from plane on the axis. Consider
the contribution to $\tau$ from a band of mass $M$ and radius $r$ around the
Galactic centre. If we assume that the band's surface density never
increases with distance from the plane, then its mass will be minimized for
a given optical depth when its surface density is constant and the
line-of-sight to the source just cuts its edge. So we take the band's
half-width to be $h(R_0-r)/R_0$, which makes the band's surface density 
 \begin{equation}
\Sigma={M\over4\pi rh}\Big(1-{r\over R_0}\Big)^{-1}.
\end{equation}
 Substituting this into equation (1) we find the band's optical depth to be
\begin{equation}\label{basicM}
\tau={GM\over c^2 h}
\end{equation}
 independent of radius (Kuijken, 1997).  This minimum mass
estimate holds if the mass is widely distributed in radius rather
than concentrated in a single band, because we can imagine a radially
continuous mass distribution to be made up of a large number of bands, and
we have shown that when the band is optimally configured, its optical depth
depends only on its mass.

From equation (\ref{basicM}) $(3.8\pm0.6)\times10^{10}\msun$ is needed
to produce the optical depth, to bulge sources that is implied by the latest
MACHO results.  A more realistic mass estimate is in excess of
$8\times10^{10}\msun$ because realistically we must assume that the surface
density of the band falls off smoothly with distance from the plane, and if
this decline is exponential with the optimal scale-height ($h[1-r/R_0]$),
the band's mass must be e times that given by (\ref{basicM}) for a given
optical depth, while if the vertical density profile is Gaussian with
optimal scale-height ($h[1-r/R_0]$), equation (\ref{basicM}) underestimates
the band's mass by a factor $\sqrt{\pi\e/2}\simeq2.07$. 

For comparison, the mass of the Galaxy interior to the Sun is of order
$M=(220\kms)^2\times8\kpc/G\simeq8.9\times10^{10}\msun$.  Thus this naivest
estimate of the mass interior to the Sun is just barely equal to the {\em
minimum\/} mass required in a circular configuration to produce the reported
optical depth, which suggests that an axisymmetric Galaxy is incompatible
with the MACHO results This must be a tentative conclusion, however, until
one has taken into account the effect on a body's circular-speed curve $v_c(r)$ of
the body being strongly flatted. Binney, Bissantz \& Gerhard (2000) show
that it {\em is\/} possible to choose the Galaxy's radial density profile in
such a way that the required optical depth is obtained without generating a
value of $v_c$ that conflicts with observation. However, such radial density
profiles require more matter in the solar neighbourhood than observations of
the Oort limit (Cr\'ez\'e et al, 1998; Holmberg \& Flynn, 2000) and the mass
within $1.1\kpc$ of the plane (Kuijken \& Gilmore, 1991) imply.
Consequently, we can safely conclude that an axisymmetric Galaxy cannot have
as large an optical depth as that reported.

Can one achieve a higher optical depth within a give mass budget by making
the bands elliptical rather than circular? Imagine deforming an initially
circular  band into an
elliptical shape while holding constant the radius $r$ at which the line of
sight to our sources cuts the band. It is straightforward to show that if
the column density through the band to the sources is to be independent of
the band's eccentricity $e$, its mass $M(e)$ must satisfy
 \begin{equation}\label{Mofe}
M(e)=M(0){1-e^2\cos^2\phi\over\sqrt{1-e^2}},
\end{equation} 
 where $\phi$ is the angle between the band's major axis and the Sun--centre
line. For $\phi<\pi/4$, $M(e)$ is a minimum with respect to $e$ at
 \begin{equation}\label{emin}
e_{\rm min}=\sqrt{2-\sec^2\phi}.
\end{equation}
 Substituting equation (\ref{emin}) in equation (\ref{Mofe}) we find the
minimum mass to be 
\begin{equation}
M_{\rm min}=M(0)\sin2\phi,
\end{equation}
 and to require axis ratio
$q_{\rm min}=\tan\phi$ (Zhao \& Mao, 1996). 

For $\phi=20\deg$, a value favoured by Binney et al.\ (1997), 
$q_{\rm min}=0.36$ and $M_{\rm min}/M(0)=0.64$; for $\phi=15\deg$, we
find $q_{\rm min}=0.27$ and $M_{\rm min}/M(0)=0.50$. Hence, making the bands
elliptical realistically reduces the mass required to generate a given
optical depth by at most $50\%$.  In practice we cannot reduce our
requirement for mass by so large a factor because the structure of the
Galaxy's stellar bar is strongly constrained by both near-IR photometry
(Blitz \& Spergel, 1991; Bissantz et al., 1997) and radio-frequency
observations of gas that flows in the Galactic plane (Englmaier \& Gerhard,
1999; Fux, 1999). Binney et al.\ (2000) estimate the possible reduction in
mass by assuming that material at $R<4\kpc$ forms a bar of optimal
eccentricity whose long axis makes an angle of $20\deg$ with the Sun--centre
line, while the Galaxy is axisymmetric at $R>4\kpc$. They show that if the
vertical structure of such a Galaxy is chosen to be optimal for lensing,
then a radial density profile can be found that nowhere exceeds the observed
value of $v_c$ and is also compatible with the constraints on the density of
matter at $R_0$. They argue, however, that such Galaxy models can be
excluded for two main reasons. First, they predict values of $v_c$ that are
too small at $R\la350\pc$ because at small $R$ they place significant mass
high above the plane, where it contributes to $\tau$ but not $v_c$. Second,
these models do not leave enough room in $v_c$ for (i) departures from the
optimal vertical profile and  (ii) the presence of  matter, such as
interstellar gas and non-baryonic matter, that  contributes to $v_c$ but not
$\tau$.

Hence, even though we don't know much about the Galactic force-field, we
know enough to exclude the measured optical depth in Baade's window! In
fact, if we were to take seriously the prediction of simulations of the
cosmological clustering of CDM that dark matter contributes substantially to
the mass interior to the Sun (e.g., Navarro \& Steinmetz, 2000), our
predicted optical depth for Baade's window would be significantly {\em
less\/} than $10^{-6}$, a factor of 4 or more below the measured value.
Something is seriously wrong here.

\end{document}